# Assessing and Improving Cybersecurity Maturity for SMEs: Standardization aspects

Bilge Yigit Ozkan✉[0000-0001-6406-356X] and Marco Spruit[0000-0002-9237-221X]

[1] Utrecht University, Department of Information and Computing Sciences, Princetonplein 5, 3584 CC Utrecht, Netherlands
`b.yigitozkan@uu.nl, m.r.spruit@uu.nl`

**Abstract.** SMEs constitute a very large part of the economy in every country and they play an important role in economic growth and social development. SMEs are frequent targets of cybersecurity attacks similar to large enterprises. However, unlike large enterprises, SMEs mostly have limited capabilities regarding cybersecurity practices. Given the increasing cybersecurity risks and the large impact that the risks may bring to the SMEs, assessing and improving the cybersecurity capabilities is crucial for SMEs for sustainability.

This research aims to provide an approach for SMEs for assessing and improving their cybersecurity capabilities by integrating key elements from existing industry standards.

## 1      Introduction

Small and Medium-sized Enterprises (SMEs) and craft enterprises are a very important part of the European economy, accounting for 99.8% of all businesses, 66.5% of all jobs and 57.6% of value added [1]. Given their large economic and social impact, there is increasing emphasis on how to serve SMEs by standards that address their peculiar characteristics.

In this paper, we investigate the requirements for a cybersecurity assessment questionnaire for SMEs to assess and improve their information security and cybersecurity capabilities by integrating key elements from existing industry standards. Our research question is "How can we meaningfully integrate existing standards for the purpose of SME self-assessment of cybersecurity maturity in a transparent way?". We are currently investigating the design of a structured method for integrating existing standards for the assessment questions.

In the background section, first, we provide information on international and European standardization initiatives with an SME perspective, in particular information security and cybersecurity related standardization. Second, we focus on information security and cybersecurity maturity models in combination with the self-assessment concept. Further, in section 3, first, we present the requirements of the model and the development of the assessment questionnaire based on standards. Second, an excerpt from the assessment questionnaire is given and the utilization of the improvement plan as a facilitator for standardization processes is discussed. In the conclusion, we discuss the scope of this research and its position in the ongoing research of designing a cybersecurity focus area maturity model for SMEs.

## 2 Background

### 2.1 International and European Standardization and SMEs

According to ISO (International Organization for Standardization), standards are documents that provide requirements, specifications, guidelines or characteristics that can be used consistently to ensure that materials, products, processes and services are fit for their purpose [2].

ISO states the benefits of standards for small to medium sized enterprises (SMEs) as follows [3]:

- Build customer confidence that your products are safe and reliable
- Meet regulation requirements, at a lower cost
- Reduce costs across all aspects of your business
- Gain market access across the world

The creation of standards derives from the experience of all interested parties who represent the meeting between the demands of society and technology harmoniously coordinated by Standardisation Bodies [4].

Standardisation is increasingly seen as a bridge between research, innovation and the market, and as a means of capturing and disseminating knowledge and, therefore, can make a positive contribution to economy, growth and prosperity at a time when Europe needs more innovation in order to remain competitive on the global stage [1].

There are three Standardisation Bodies on European level:

- CEN (European Committee for Standardization)
- CENELEC (European Committee for Electrotechnical Standardization)
- ETSI (European Telecommunications Standards Institute)

There is also the Small Business Standards (SBS) association to represent SMEs in the standardization process in Europe. SBS is a European non-profit association co-financed by the European Commission and EFTA Member States. Its goal is to represent and defend SMEs' interests in the standardisation process at European and international levels. Moreover, it aims at raising the awareness of SMEs about the benefits of standards and at encouraging them to get involved in the standardisation process [5].

### 2.2 Information Security Standards and SMEs

Since, data breaches and cyber-attacks are becoming a regular occurrence, ISO 27001:2013 the information security management system standard [6] is adapted by a large number of organizations all over the globe. According to an ISO Survey [7], in 2016, 33,290 certifications were issued worldwide, compared to 27,536 certifications the previous year.

Recently, Small Business Standards (SBS) published an SME Guide for the implementation of ISO/IEC 27001 on information security management to help with establishing or raising information security levels within an SME [8]. ISO 27002 [9] is a standard designed to be used by organizations that intend to:

- select controls within the process of implementing an Information Security Management System based on ISO/IEC 27001;
- implement commonly accepted information security controls;
- develop their own information security management guidelines.

We opted to investigate mostly ISO 27001 and ISO 27002 to drive the questions for the assessment of capabilities due to their comprehensiveness and prominent position in the domain of information/cyber security. A holistic study funded by Dutch Government for the inventorization and the classification of existing cybersecurity standards, presents the comprehensive coverage of the ISO 27001 standard in several dimensions [10].

### 2.3 Information Security and Cybersecurity Maturity Models

Maturity Modelling is a method for representing domain specific knowledge in a structured way in order to provide organizations with an evolutionary process for assessment and improvement. Maturity models in different domains have been developed and used mostly since they became popular after the introduction of the Capability Maturity Model (CMM) of the Software Engineering Institute (SEI) of Carnegie Mellon University [11]. There is an abundance of work related to information security and cybersecurity maturity modelling. Some of these maturity models are given in Table 1.

**Table 1.** Information and Cybersecurity Maturity Models

| Maturity Model | Organization/ Authors |
| --- | --- |
| Cybersecurity Capability Maturity Model[12] | US The Department of Energy (DOE) |
| Open Information Security Management Maturity Model[13] | The Open Group |

| NICE Cybersecurity Capability Maturity Model[14] | US The Department of Homeland Security |
| --- | --- |
| ISFAM (the Information Security Focus Area Maturity Model)[15] | Spruit & Roeling |

Spruit & Roeling [15] developed the Information Security Focus Area Maturity Model (ISFAM) which is capable of determining the current information security maturity level and can be used to incrementally and structurally improve information security maturity within the organization. ISFAM is successfully evaluated through several case studies in telecommunications, logistics, healthcare and finance sectors.

The focus area type maturity model was first proposed in 1999 [16]. Design of this type of maturity models is then notably formalized in detail. The assessment questionnaire development is defined as a step called "Develop assessment instrument" in the design process of focus area maturity models which is further explained in section 3.2[17].

### 2.4  Self-assessment

As the term implies, self-assessment is a means by which an organization assesses compliance to a selected reference model or module without requiring a formal method [18]. The information/cyber security maturity models given in Table 1 are complex and comprehensive. They are not easy to implement for self-assessment and suitable for preparing customized improvement plans.

## 3  Towards a Standards-Transparent Focus Area Maturity Model for Assessing and Improving Cybersecurity for SMEs

### 3.1  The Requirements for the Model

Previous research [15, 19] pointed out the following requirements to be met in the process of developing a model suitable for assessing and improving cybersecurity for SMEs.

- **Easy to use, self-assessment, do-it-yourself**

Assessment and improvement planning should be easy to be utilized by SMEs, requiring minimal extra resources.

- **Situational awareness**

Different characteristics of an organization/entity should be considered while designing adaptive maturity assessment models. We will further investigate the applicability of the CHOISS model [19] for designing adaptive maturity assessment models. The model should be able to provide a customized guidance and implementation plan.

- **Standards-transparency**

The model should support the capability of adhering to related standards on cybersecurity. The relation between the cybersecurity capabilities and related standards should be transparent.

- **Provide cybersecurity awareness**

The model should provide training material and increase awareness on cybersecurity with regard to the assessed capabilities.

- **Maintainability by design**

Given the ever-changing and dynamic nature of the cybersecurity threats, the model should be capable of incorporating new capabilities and standards.

Currently, we are conducting research that aims to handle abovementioned requirements. In this paper, we focus on the standards-transparency requirement for the model.

### 3.2 Developing the Assessment Questionnaire for the Model based on Standards

To be able to use a focus area maturity model as an instrument to assess the current maturity of a functional domain, measures must be defined for each of the capabilities. This can be done by formulating control questions for each capability. These questions can be combined in a questionnaire that can be used in assessments. Formulation of the questions is usually based on the descriptions of the capabilities and on experience and practices [17].

In this research, we have used ISO 27001 [6], ISO 27002 [9] and ETSI TR 103 305 [20] to identify the assessment questions for the 'Identity Management and Access Control' focus area. In Table 2, the referenced standard and the specific clause is given in the third column. Along with the standards, literature can also be used for identifying the required capabilities and the assessment questions.

### 3.3 A Standard-Transparent Assessment Questionnaire for Identity Management and Access Control

In a focus area maturity model *Capability* is defined as the ability to achieve an objective for the focus area [17]. Each focus area consists of a number of different capabilities representing progressive maturity levels.

In the assessment questionnaire, questions are grouped by capability level (A, B and C). Table 2 shows some example questions selected for the Identity Management and Access Control focus area for each capability level.

**Table 2.** An excerpt from the Identity Management and Access Control Questionnaire

| Question Number | Question | Standard, Clause | Capability Level | Question Type |
|---|---|---|---|---|
| F1Q1 | Do your users login to your systems by unique user-ids? | ISO 27002, 9.2.1.a | A | Scale |
| F1Q2 | Do you periodically review your access rights (including administrator accounts)? | ISO 27002, 9.2.2.f 9.2.3.f 9.2.5, ETSI TR 103 305, CSC 16 | B | Scale |
| F1Q3 | How frequently do you review your access rights (including administrator accounts)? | ISO 27002, 9.2.5 | B | Multiple choice |
| F1Q4 | When have you reviewed your access rights (including administrator accounts) the last time? | - | B | Date/ Time |
| F1Q5 | Do you implement segregation of access control roles, e.g. access request, access authorization, and access administration? | ISO 27002, 9.2.2.b, 6.1.2 | C | Scale |

Questions are answered by the SME using the following implementation level scale given in Table 3. In this table the score contribution percentage per implementation level is also given. In our model this ratio is used to calculate the score achieved for a given capability level.

**Table 3.** Implementation Level Scale

| Implementation Level | % Contribution to the Score |
|---|---|
| Fully Implemented (FI) | 100 |
| Largely Implemented (LI) | 85 |
| Partially Implemented (PI) | 50 |
| Not Implemented (NI) | 0 |

When the SME performs self-assessment using the assessment questionnaire, based on the given answers, the capabilities that are not currently implemented can be used to form a customised improvement plan that also facilitates the standardization efforts.

The standards transparency also supports SMEs to have quick reference for the capabilities and the focus areas and increase their standards awareness. In Table 4, the capability improvement plan for an SME who has not implemented the first two capabilities yet is given as an example.

**Table 4.** An Exemplar Capability Improvement Plan

| Information Security Capability Improvement Plan For UU | | | |
|---|---|---|---|
| **Identity Management and Access Control Tasks** | | **Capability Score : 50%** | |
| Task Number | Description | Deadline | Responsible |
| T1 | Ensure that users login to the systems by unique user-ids. | 01/08/2018 | B.Y. Ozkan |
| T2 | Ensure that access rights (including administrator accounts) are periodically reviewed. | 01/08/2018 | B.Y. Ozkan |

## 4       Conclusion

Using the proposed self-assessment questionnaire, SMEs can assess their information security and cybersecurity capabilities and identify areas of improvement in a standards- transparent way. The assessment questionnaire can also assist SMEs in creating a plan to improve the cybersecurity practices, thereby reaching a higher maturity level.

The actual application and implementation of the assessment questionnaire within an organization is beyond the scope of this research. Further research has to be conducted to validate the questionnaire model and to monitor its applicability. In the SMESEC project, evaluation of the proposed assessment questionnaire to assess and improve the cybersecurity capabilities of SMEs is planned. In this paper, we have only focused on standards-transparency aspects of the model requirements. Our ongoing research in the SMESEC project is focused on all the requirements presented herein.

## Acknowledgements

This work was made possible with funding from the European Union's Horizon 2020 research and innovation programme under grant agreement No 740787 (SMESEC). The opinions expressed and arguments employed herein do not necessarily reflect the official views of the funding body.